# Acetylene weak bands at 2.5 µm from intracavity Cr²⁺:ZnSe laser absorption observed with time-resolved Fourier transform spectroscopy.

**Véronique Girard [a], Robert Farrenq [a], Evgeni Sorokin [b], Irina T. Sorokina [b], Guy Guelachvili [a], and Nathalie Picqué [a]**

[a] Laboratoire de Photophysique Moléculaire, Unité Propre du C.N.R.S., Bâtiment 350, Université de Paris-Sud, 91405 Orsay, France
[b] Institut für Photonik, TU Wien, Gusshausstr. 27/387, A-1040 Vienna, Austria

Corresponding author:
Dr. Nathalie Picqué,
Laboratoire de Photophysique Moléculaire
Unité Propre du CNRS, Université de Paris-Sud, Bâtiment 350
91405 Orsay Cedex, France
Website: http://www.laser-fts.org
Phone nb: 33 1 69 15 66 49
Fax nb: 33 1 69 15 75 30
Email: nathalie.picque@ppm.u-psud.fr

Abstract

The spectral dynamics of a mid-infrared multimode Cr²⁺:ZnSe laser located in a vacuum sealed chamber containing acetylene at low pressure is analyzed by a stepping-mode high-resolution time-resolved Fourier transform interferometer. Doppler-limited absorption spectra of $C_2H_2$ in natural isotopic abundance are recorded around 4000 cm⁻¹ with kilometric absorption path lengths and sensitivities better than $3 \ 10^{-8}$ cm⁻¹. Two cold bands are newly identified and assigned to the $\nu_1+\nu_4^1$ and $\nu_3+\nu_5^1$ transitions of $^{12}C^{13}CH_2$. The $\nu_1+\nu_5^1$ band of $^{12}C_2HD$ and fourteen $^{12}C_2H_2$ bands are observed, among which for the first time $\nu_2+2\nu_4^2+\nu_5^{-1}$.

1. Introduction

    The infrared spectroscopy of the acetylene molecule $C_2H_2$ is of interest for theoretical, atmospheric, planetary, astrophysical, metrological and industrial applications [1]. In particular, experimental laboratory detection of weak lines from $C_2H_2$ hot bands can help in improving global treatment of acetylene in its ground state using effective Hamiltonian approaches and in modeling the dense spectra from circumstellar envelopes or giant planets atmosphere.

    The 2.5 µm domain is the seat of perpendicular bands of $^{12}C_2H_2$ and its isotopologues. Strong $Q$ branches may make it interesting for applications linked to its detection in various media. However, the region has not been extensively investigated yet. The strong $v_3+v_4^1$ and $v_2+2v_4^0+v_5^1$ bands of $^{12}C_2H_2$, respectively centered at 3897 and 3881 $cm^{-1}$, have been observed by Palmer et al [2] and by Lafferty and Pine [3]. Two cold and five hot bands from absorption Doppler-limited $^{12}C_2H_2$ spectra with pressure-path length products at most equal to 96 Torr.m, have been analyzed by D'Cunha et al [4]. As far as the other isotopologues are concerned, only a cold band and three hot bands from $^{12}C_2HD$ have been reported [5].

    Intracavity laser absorption is a powerful approach to the spectroscopy of weak transitions. The absorbing sample is placed inside a broad gain bandwidth laser cavity. Broadband losses are compensated by the laser gain unlike spectrally selective losses mostly due to the sample absorption lines. Most often, the multimode laser is pulsed mode operated. Then, absorption is expected to follow the Lambert-Beer law with path length $L = c\,t_g$, where $c$ is the velocity of light and $t_g$ is the generation time, separating the population inversion threshold from the observation. Equivalent absorption path length of several tens of kilometers can be achieved, resulting in high detection sensitivities. Spectra cover several tens of reciprocal centimeters.

    Recently we reported [6] implementation of high resolution Time-Resolved Fourier Transform IntraCavity Laser Absorption Spectroscopy (called hereafter TRFT-ICLAS). Coupling a $Cr^{2+}$:ZnSe laser with a stepping mode interferometer, we were able [7] to reach the 2.5 µm region, extreme infrared limit ever reached by ICLAS with Doppler-limited resolution. Since, improved experimental setup enabled to record new spectra. In this letter, we briefly report the latest implementation, describe the recorded spectra and present the analysis of weak acetylene bands between 3910 and 4030 $cm^{-1}$.

2. Experimental

    The experimental set-up is schematized on Figure 1. A multimode $Cr^{2+}$:ZnSe laser is optically pumped by a 1607 nm $Er^{3+}$-fiber laser. An acousto-optic modulator chops the pumping power above and below the $Cr^{2+}$:ZnSe threshold in order to initiate its spectro-temporal dynamics. The pump is focused on a 4-mm long $Cr^{2+}$:ZnSe Brewster-cut crystal. The $Cr^{2+}$:ZnSe laser comprises a four-mirror astigmatism-compensated X-shaped resonator. $M_1$ and $M_2$ are highly reflective (HR) spherical mirrors with a reflection coefficient equal to 99.8% between 3700 and 4300 $cm^{-1}$ and a radius of curvature equal to 75 mm and 100 mm respectively. $M_3$ and output coupler OC are HR plane mirrors, with the same coating as $M_1$ and $M_2$. The small portion (T<0.2%) of light leaking outside OC is sent to the stepping-mode interferometer. Use of an OC with a much smaller transmission than in [7] (0.2 % instead of 6%) dramatically improves the laser dynamics and enables higher absorption path lengths excursions. The threshold under cw operation is as low as 37 mW. No wavelength selective element is inserted inside the cavity and consequently no attempt to tune the laser is made. The $Cr^{2+}$:ZnSe cavity is inside a vacuum enclosure. In order to get rid of strong atmospheric absorption from water vapor, the laser is first evacuated under secondary vacuum conditions (of the order of $10^{-2}$ Pa $=10^{-4}$ mbar) and then filled with the $C_2H_2$ gas. This procedure with no need of insertion of a gas cell in the cavity provides the benefits of optimal cavity filling ratio

and of no restriction of the spectral domain width otherwise degraded by additional losses due to the cell windows. The interferometer is operated under primary vacuum. It is equipped with a $CaF_2$ beamsplitter and two liquid nitrogen cooled InSb detectors. The rise time of these detectors is about 0.5 µs and this represents the actual time resolution limitation. The TRFT-ICLAS data acquisition procedure has been detailed elsewhere [6]. Briefly, the whole acquisition procedure is computer-driven. At given path difference step, the time sampling process is triggered by an InGaAs photodiode receiving a small portion of the pumping beam. In order to improve dynamic range of the measurement, detector signals delivered at two outputs of the Connes-type interferometer are balanced and subtracted. This removes the dc component from the interferogram [8]. An analog-to-digital-converter acquisition board (500-MHz, 8 bits) digitalizes this interferometric signal. A digital input/output device synchronizes data acquisition and interferometer step-scan procedure. Several laser shots may be co-added in order to improve the signal to noise ratio. The same time-sampling is repeated at each path difference step, up to maximum path difference. Finally as many interferograms as time-samples per step are collected. Each transformed interferogram is a complete absorption spectrum at a given generation time or in other words, thanks to ICLAS with a given equivalent absorption length. The pulse-to-pulse variability and frequency drifts of the optimized laser have been especially looked at and as they are found to be negligible, they are consequently not taken care of during the recording procedure.

Five high-resolution acetylene time-resolved intracavity absorption spectra have been recorded, with different pressures. The recording conditions are summarized in Table 1. A natural abundance acetylene sample (L'Air Liquide, Purity>99.6%) is used at room temperature (296 ± 0.5 K) and its pressure is monitored with a Baratron gauge. Figure 2 provides an intermediate-resolution ($0.07 \text{ cm}^{-1}$) representation of the time-resolved spectrum nb. 505. For clarity, only one time-component out of 5 is displayed, from time-component nb.5 to nb.60, without normalization. Their maximum intensity evolves like the total laser intensity shown on Figure 3. Actually, the spectrum, which results from a 252 minutes experiment, has 64 time-components. Each time-component is made of 80,000 independent spectral samples. The time component nb. 5 (23.2 µs corresponding to $L = 7.0$ km), extends on $120 \text{ cm}^{-1}$ centered at $3965 \text{ cm}^{-1}$. At $9.2 \ 10^{-3} \text{ cm}^{-1}$ full unapodized resolution, its signal to noise ratio is at most 50. Its corresponding minimum detectable absorption coefficient is of the order of $2.9 \ 10^{-8} \text{ cm}^{-1}$. As expected, the envelope of the spectrum gets sharper and more intense with increasing generation times. The time-component nb. 60, (111.2 µs, corresponding to $L = 33.3$ km) is about $50 \text{ cm}^{-1}$ broad. With the $0.07 \text{ cm}^{-1}$ intermediate resolution shown on the figure, the minimum detectable absorption coefficient of this time-component is about $3.5 \ 10^{-9} \text{ cm}^{-1}$. Strong fringes originating from intracavity etalon effect appear at higher resolution and increase with generation time. Doppler-limited long generation times time-components consequently appear with a degraded signal-to-noise ratio. No attempt was undertaken to suppress them and the $C_2H_2$ analysis was made on Doppler-limited time-components with generation time uppermost values of the order of 30 µs. The total intensity variation of the laser beam is given on Figure 3. The evolution versus the time of the peak absorbance of five acetylene lines is also shown on the figure. The absorbance is simply defined as $\ln(1-I_a/I_0)$ with $I_a$ absorption at line center and $I_0$ local background. The linearity of the evolution demonstrates the validity of the equation $L = c \ t_g$ in the corresponding time domain ( $t_g < 110$ µs ).

## 3. Analysis

Recorded spectra are highly congested and exhibit essentially lines never measured or even predicted before [2-5, 9-11]. The wavenumber scales are calibrated using residual water vapor lines, with data from [12]. The accuracy of the acetylene line positions is of the order of $10^{-3}$ cm$^{-1}$. Assignments reported in this paper were greatly facilitated by the use of Loomis–Wood programs [13,14]. However about 55 % of the measured lines remain unassigned, among which many are obviously part of polynomial series.

Identified bands are given in Table 2. The $\nu_2+2\nu_4{}^2+\nu_5{}^{-1}$ band of $^{12}C_2H_2$ is newly detected. It was recognized with predictions and assignments from the effective Hamiltonian developed by [15, 16]. The remaining thirteen $^{12}C_2H_2$ bands have already been observed by [2,4,11] and their lines reported here extend previous measurements towards high $J$ values. The $^{12}C_2HD$ $\nu_1+\nu_5{}^1$ band has already been observed [5]. Two bands with strong $Q$ branches, respectively centered at 3953.7 and 4005.0 cm$^{-1}$, are also present in the spectra. They could not be assigned using constants from observed or predicted vibrational levels, of any acetylene isotopologue, available in the literature, especially from [9,10]. We have assigned these two bands respectively to the $\nu_1+\nu_4{}^1$ and $\nu_3+\nu_5{}^1$ transitions of $^{12}C^{13}CH_2$, using predictions of the upper state levels based on isotopic shifts. Their absorption is consistent with the abundance of $^{12}C^{13}CH_2$ (2.197 %) in a natural sample and the expected intensity of a cold band. These attributions have been confirmed by the retrieved principal rotational constants $B_v$ and their comparison with the calculated values from the spectroscopic parameters given in [10,17]. Figure 4 displays a small high-resolution portion of time-component nb.5 of spectrum 505 showing the $Q$-branch of the $\nu_1+\nu_4{}^1$ band of $^{12}C^{13}CH_2$.

Fitting all experimental line positions of a given band provides the effective rotational constants. The energy $E$ of the rotational level of quantum number $J$ from the vibrational level $\upsilon_1\upsilon_2\upsilon_3\left(\upsilon_4{}^{\ell_4}\upsilon_5{}^{\ell_5}\right)$ is expressed as:

$$E=G_v\left(\upsilon_1\upsilon_2\upsilon_3\left(\upsilon_4{}^{\ell_4}\upsilon_5{}^{\ell_5}\right)\right)+B_v\left(J(J+1)-\left(\ell_4+\ell_5\right)^2\right)-D_v\left(J(J+1)-\left(\ell_4+\ell_5\right)^2\right)^2+H_v\left(J(J+1)-\left(\ell_4+\ell_5\right)^2\right)^3$$

where $\ell_4$ and $\ell_5$ are the bending vibrational angular momenta. The standard deviation of the fits are given in Table 2. Table 3 reports the calculated constants of the levels obtained from the fits of the $\nu_2+2\nu_4{}^2+\nu_5{}^{-1}$ transition of $^{12}C_2H_2$ and of the transitions belonging to the $^{12}C_2HD$ and $^{12}C^{13}CH_2$ acetylene isotopologues. The lower state constants in the table are held fixed in the fitting procedure and are taken in Ref.[5,10,18]. $^{12}C_2H_2$ effective constants from the already observed bands are not given since wavenumbers of the higher $J$ value observed transitions strongly deviate from a polynomial model due to perturbations.

A list of observed line positions of the seventeen bands with assignments and (obs.-calc.) values is attached to this paper as supplementary material. Comparison with the experimental line positions from the bands reported by [2-4] is difficult because we only share the observation of a few common high $J$ lines, which are often saturated and blended in our spectra. Selected isolated lines give an agreement better than $10^{-3}$ cm$^{-1}$, but the overall agreement is within $\pm$ 5 $10^{-3}$ cm$^{-1}$. There is no overlapping between the rotational lines of the $\nu_3+2\nu_4{}^2-\nu_4{}^1$, $\nu_3+(\nu_4+\nu_5)^0{}_+-\nu_5{}^1$, $\nu_3+(\nu_4+\nu_5)^0{}_--\nu_5{}^1$, $\nu_3+(\nu_4+\nu_5)^2-\nu_5{}^1$ bands reported here and those from [11]. Reasonable agreement is obtained with the predictions provided by [16] : the lines correspond within $\pm$ 2 $10^{-2}$ cm$^{-1}$ and the discrepancies increase with $J$.

4. Conclusion

A new experimental set-up, based on time-resolved Fourier transform intracavity Cr$^{2+}$:ZnSe laser absorption and dedicated to high sensitivity spectroscopy in the 2.5 µm mid-infrared region, has been applied to the spectroscopy of $C_2H_2$. Line positions of seventeen vibrational bands, including two new transitions, of acetylene are reported in a spectral region where laboratory spectra are missing. The strong percentage of lines which remain unassigned shows however that broader experimental spectral coverage and refined models including most abundant isotopologues transitions are needed.

Acknowledgments

V.I. Perevalov is warmly acknowledged for providing his latest effective Hamiltonian $^{12}C_2H_2$ line position predictions. We also thank J.-Y. Mandin for useful discussions and comments. Participation to the experiments from F. Gueye and H. Herbin was appreciated. This research has been supported by the French-Austrian Amadeus exchange program.

Table captions

Table 1:  Recording conditions for the high resolution $C_2H_2$ absorption time-resolved spectra.

Table 2: Bands of $^{12}C_2H_2$, $^{12}C_2HD$ and $^{12}C^{13}CH_2$ assigned in the present TRFT-ICLAS spectra between 3910 and 4030 $cm^{-1}$. The laser emission spectral domain most often restricts the observation to only part of one branch of the band. The band center is equal to $G'_v - B'_v k'^2 - D'_v k'^4 - ( G''_v - B''_v k''^2 - D''_v k''^4 )$, where $k=\ell_4+\ell_5$. Numbers in parentheses are one standard deviation in units of the least significant digits.

Table 3 : Effective molecular constants (in $cm^{-1}$) of the levels obtained from the fits of the transitions belonging to the $^{12}C_2HD$ and $^{12}C^{13}CH_2$ acetylene isotopologues. The lower state constants were held fixed in the fits. Numbers in parentheses are one standard deviation in units of the least significant digits.

| Spectrum number | Pressure (Pa) | Unapodized spectral resolution ($10^{-3}$ cm$^{-1}$) | Number of time samples | Time resolution (µs) | Number of coadditions | Pump power (mW) | Pumping ratio (P/P$_{threshold}$) | Experiment duration (minutes) |
|---|---|---|---|---|---|---|---|---|
| 498 | 7 | 10.5 | 64 | 1.6 | 32 | 70 | 1.87 | 35 |
| 499 | 75 | 10.5 | 64 | 1.6 | 32 | 70 | 1.87 | 35 |
| 500 | 2720 | 9.2 | 64 | 1.6 | 32 | 70 | 1.87 | 44 |
| 501 | 673 | 73.7 | 256 | 3.2 | 16 | 125 | 1.95 | 9 |
| 505 | 673 | 9.2 | 64 | 1.6 | 256 | 50 | 1.34 | 252 |

Table 1

| Isotopologue | Transition | Band type | Observed lines* | Band center $G_c$ (cm$^{-1}$) | RMS fit ($10^{-3}$ cm$^{-1}$) | Already reported lines* |
|---|---|---|---|---|---|---|
| $^{12}C_2H_2$ | $0011^10^0 - 0000^00^0$ | $\Pi_u \leftarrow \Sigma_g^+$ | $R_{ee}$ (24-41) | 3897.43(2) | 3.81 | $P_{ee}$(0-36) $R_{ee}$(0-37) [2] |
| | $0011^11^{-1} - 0000^01^1$ | $\Sigma_g^+ \leftarrow \Pi_u$ | $R_{ee}$ (21-25) | 3880.186(7) | 1.65 | $P_{ee}$(1-19) $Q_{ef}$(1-19) $R_{ee}$(1-18) [11] |
| | $0011^{-1}1^1 - 0000^01^1$ | $\Sigma_g^- \leftarrow \Pi_u$ | $R_{ff}$ (18-24) | 3888.7714(6) | 1.03 | $P_{ff}$(1-16) $Q_{fe}$(1-20) $R_{ff}$(1-18) [11] |
| | $0011^11^1 - 0000^01^1$ | $\Delta_g \leftarrow \Pi_u$ | $R_{ee}$ (18-30) | 3890.904(1) | 2.18 | $P_{ee}$(3-19) $Q_{ef}$(2-21) $R_{ee}$(1-20) [11] |
| | | | $R_{ff}$ (18-37) | 3890.9030(8) | 1.67 | $P_{ff}$(3-23) $Q_{ef}$(2-21) $R_{ff}$(1-23) [11] |
| | $0012^00^0 - 0001^10^0$ | $\Sigma_u^+ \leftarrow \Pi_g$ | $R_{ee}$ (14-24) | 3898.665(1) | 3.31 | $P_{ee}$(1-22) $Q_{ef}$(1-20) $R_{ee}$(1-19) [11] |
| | $0012^20^0 - 0001^10^0$ | $\Delta_u \leftarrow \Pi_g$ | $R_{ee}$ (22-30) | 3892.871(6) | 12.9 | $P_{ee}$(3-25) $Q_{ef}$(2-22) $R_{ee}$(1-24) [11] |
| | | | $R_{ff}$ (26-36) | 3892.8417(7) | 1.83 | $P_{ff}$(3-28) $Q_{ef}$(2-19) $R_{ff}$(1-27) [11] |
| | $0102^01^1 - 0000^00^0$ | $\Pi_u \leftarrow \Sigma_g^+$ | $R_{ee}$ (28-43) | 3881.34(2) | 2.34 | $P_{ee}$(2-39) $Q_{ef}$(9-35) $R_{ee}$(0-37) [2] |
| | $0102^21^{-1} - 0000^00^0$ | $\Pi_u \leftarrow \Sigma_g^+$ | $R_{ee}$(11-29) | 3906.986(5) | 1.47 | - |
| | $1000^01^1 - 0000^00^0$ | $\Pi_u \leftarrow \Sigma_g^+$ | $P_{ee}$ (33-47) | 4091.17(1) | 0.73 | $P_{ee}$(2-35) $Q_{ef}$(1-35) $R_{ee}$(0-33) [4] |
| | $1000^02^0 - 0000^01^1$ | $\Sigma_g^+ \leftarrow \Pi_u$ | $P_{ee}$ (26-42) | 4071.85(3) | 7.42 | $P_{ee}$(1-30) $Q_{ef}$(1-27) $R_{ee}$(1-30)[4] |
| | $1000^02^2 - 0000^01^1$ | $\Delta_g \leftarrow \Pi_u$ | $P_{ee}$ (25-34) | 4080.08(2) | 1.18 | $P_{ee}$(3-28) $Q_{ef}$(2-30) $R_{ee}$(1-26) [4] |
| | | | $P_{ff}$ (24-34) | 4080.29(3) | 3.12 | $P_{ff}$(3-27) $Q_{ef}$(2-27) $R_{ff}$(1-30) [4] |
| | $1001^11^{-1} - 0001^10^0$ | $\Sigma_u^+ \leftarrow \Pi_g$ | $P_{ee}$ (22-39) | 4062.23(3) | 9.4 | $P_{ee}$(1-30) $Q_{ef}$(1-28) $R_{ee}$(1-29) [4] |
| | $1001^{-1}1^1 - 0001^10^0$ | $\Sigma_u^- \leftarrow \Pi_g$ | $P_{ff}$ (26-33) | 4075.69(7) | 0.96 | $P_{ff}$(1-30) $Q_{fe}$(1-26) $R_{ff}$(1-26) [4] |
| | $1001^11^1 - 0001^10^0$ | $\Delta_u \leftarrow \Pi_g$ | $P_{ff}$ (24-37) | 4077.16(4) | 0.80 | $P_{ff}$(3-30) $Q_{ef}$(2-30) $R_{ff}$(1-27) [4] |
| $^{12}C_2HD$ | $1000^01^1 - 0000^00^0$ | $\Pi \leftarrow \Sigma^+$ | $P_{ee}$ (1-21), $R_{ee}$ (0-10) | 3995.7505(4) | 1.14 | not given [5] |
| | | $\Pi \leftarrow \Sigma^+$ | $Q_{fe}$ (6-27) | 3995.7469(4) | 1.14 | not given [5] |
| $^{12}C^{13}CH_2$ | $0010^01^1 - 0000^00^0$ | $\Pi \leftarrow \Sigma^+$ | $P_{ee}$ (2-22), $R_{ee}$ (0-8) | 4005.0495(8) | 1.76 | - |
| | | $\Pi \leftarrow \Sigma^+$ | $Q_{fe}$ (7-21) | 4005.0452(8) | 1.76 | - |
| | $1001^10^0 - 0000^00^0$ | $\Pi \leftarrow \Sigma^+$ | $P_{ee}$ (3-8), $R_{ee}$ (0-17) | 3953.6735(8) | 1.32 | - |
| | | $\Pi \leftarrow \Sigma^+$ | $Q_{fe}$ (6-22) | 3953.6681(8) | 1.32 | - |

*) Between parentheses, $J_{min}$-$J_{max}$

Table 2

| Isotopologue | State | $e/f$ | $G_v$ | $B_v$ | $D_v$ x $10^6$ | $H_v$ x $10^9$ | Ref. |
|---|---|---|---|---|---|---|---|
| $^{12}C_2H_2$ | $0000^00^0$ | $e$ | 0 | 1.176646 | 1.62710 | - | [18] |
| | $0102^21^{-1}$ | $e$ | 3908.1614(54) | 1.175848(38) | 3.214(77) | 0.888(46) | This work |
| $^{12}C_2HD$ | $0000^00^0$ | $e$ | 0 | 0.99153 | 1.120 | - | [5] |
| | $1000^01^1$ | $e$ | 3996.73688(38) | 0.9863894(63) | 1.079(17) | - | This work |
| | | $f$ | 3996.73688(38) | 0.9899973(37) | 1.1533(57) | - | This work |
| $^{12}C^{13}CH_2$ | $0000^00^0$ | $e$ | 0 | 1.148460772 | 1.556658 | - | [10] |
| | $0010^01^1$ | $e$ | 4006.19221(83) | 1.142697(11) | 1.490(23) | - | This work |
| | | $f$ | 4006.19221(83) | 1.146954(11) | 1.240(26) | - | This work |
| | $1001^10^0$ | $e$ | 3954.81431(79) | 1.140828(14) | 1.464(50) | - | This work |
| | | $f$ | 3954.81431(79) | 1.1462291(86) | 1.797(17) | - | This work |

Table 3

Figure Captions

Figure 1: Schematic representation of the TRFT-ICLAS experimental set-up. An Er-doped fiber laser is chopped by an acousto-optic modulator (AOM) and is focused by a lens (L) on the $Cr^{2+}$:ZnSe crystal for optical pumping. A portion of the pumping beam is collected by a photodetector (PD) which triggers the data acquisition. The $Cr^{2+}$:ZnSe laser resonator is made of the highly-reflective spherical $M_1$, $M_2$ and plane $M_3$ mirrors and OC output coupler. The dashed rectangle represents the vacuum-tight chamber. The light leaking out the cavity is analyzed by a high resolution Fourier spectrometer and is detected by two InSb photodetectors (PD). A personal computer (PC) drives the whole experimental procedure.

Figure 2: $C_2H_2$ time-resolved spectrum made of 64 time-components. Only one time-component out of five has been plotted, starting from time-component nb.5. On the plot, two consecutive components, at an intermediate 0.07 $cm^{-1}$ apodized resolution, are 8 µs from each other. This corresponds to a 2.4-kilometer increase of the equivalent absorbing path $L$.

Figure 3: Experimental values of the peak absorbance for five acetylene lines together with the dependence of the total laser intensity versus time. Time origin is set when optical pumping is switched on. Generation time actually begins when population inversion reaches its threshold. For a transition-metal ion doped solid-state laser such as $Cr^{2+}$:ZnSe, this may take several microseconds. Consequently peak absorbance value appears not to be zero at zero time.

Figure 4: High resolution portion of time-component nb.5 of TRFT-ICLAS spectrum nb.505 showing the $Q$ branch of the band $\nu_1 + \nu_4{}^1$ of $^{12}C^{13}CH_2$. Acetylene pressure is 673 Pa. Absorption path length is 7.0 km ($t_g$=23.2 µs). On the plot, the spectrum is apodized (resolution: 18.4 $10^{-3}$ $cm^{-1}$) Stars indicate residual water lines.

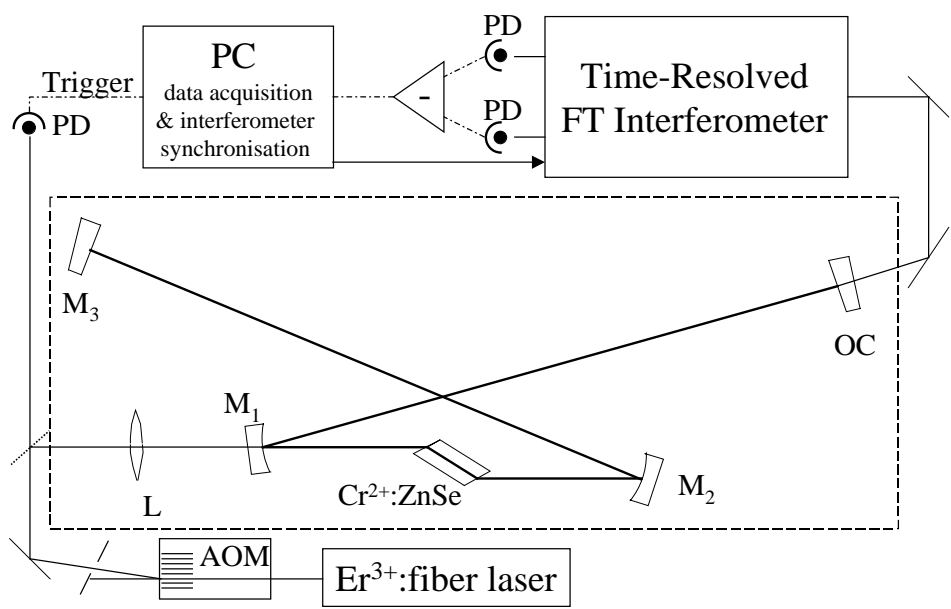

Figure 1

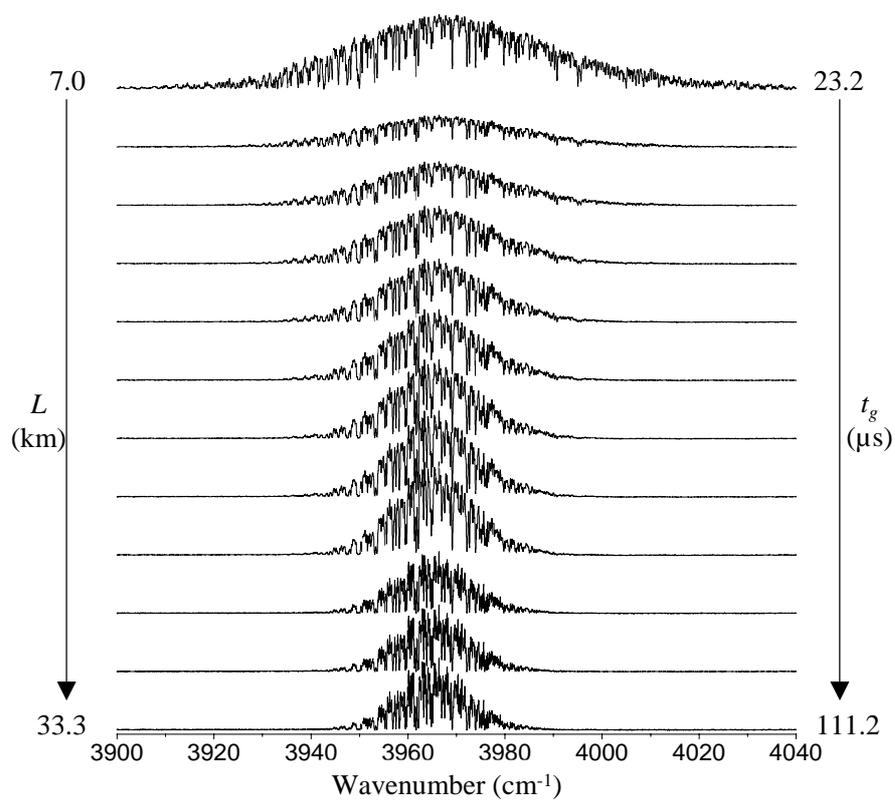

Figure 2

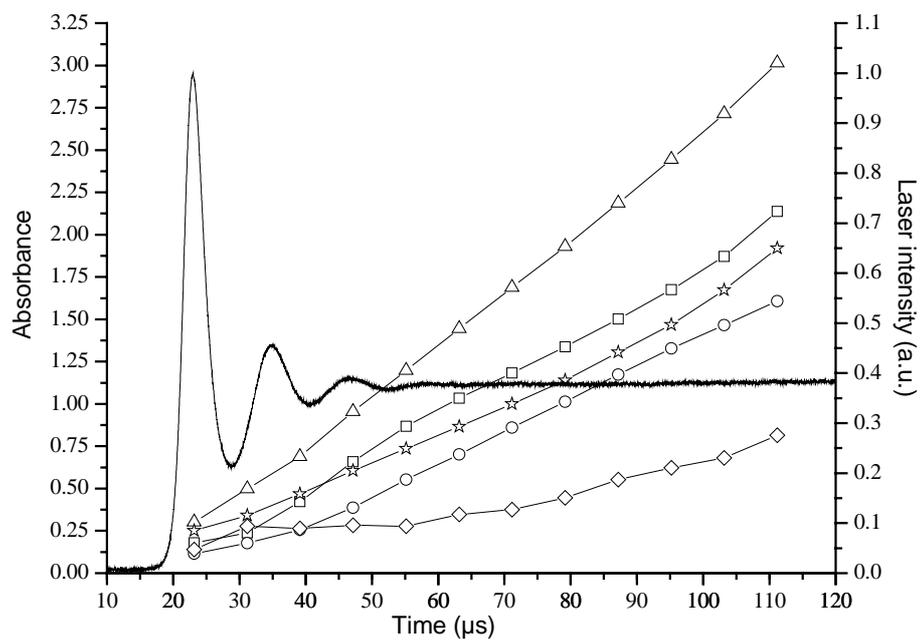

Figure 3

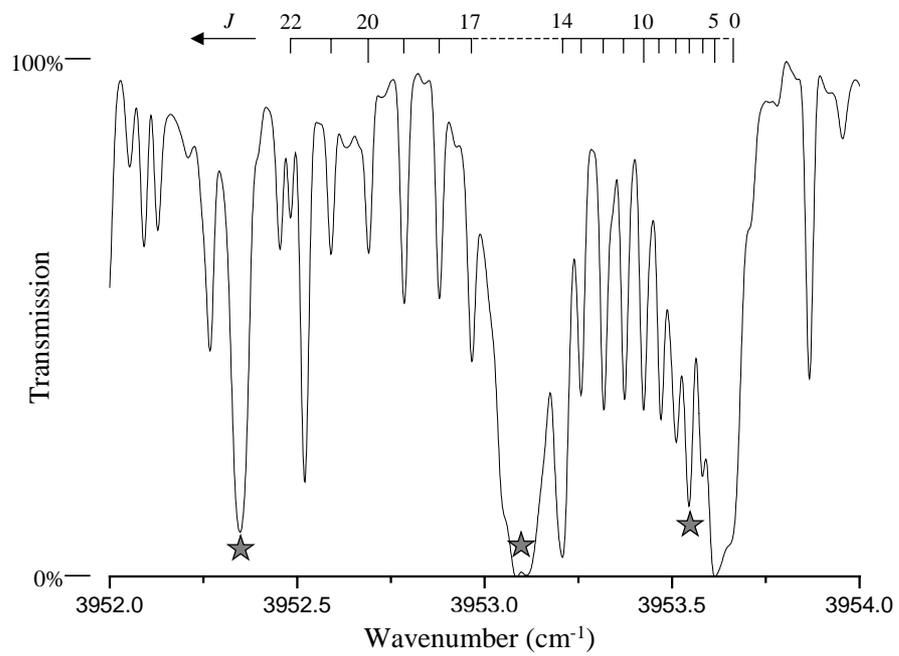

Figure 4